\documentclass[
	a4paper, 
	10pt, 
	unnumberedsections, 
	twoside, 
]{LTJournalArticle}

\usepackage{indentfirst}

\usepackage{amsmath}
\newtheorem{theorem}{THEOREM}

\title{Defending against Data Poisoning Attacks in Federated Learning via User Elimination} 

\author{%
    	Nick Galanis\textsuperscript{1}
}

\date{\footnotesize\textsuperscript{\textbf{1}}School of Engineering, University College London\\  }

\renewcommand{\maketitlehookd}{%
	\begin{abstract}
		\noindent In the evolving landscape of Federated Learning (FL), a new type of attacks concerns the research community, namely Data Poisoning Attacks, which threaten the model integrity by maliciously altering training data. This paper introduces a novel defensive framework focused on the strategic elimination of adversarial users within a federated model. We detect those anomalies in the aggregation phase of the Federated Algorithm, by integrating metadata gathered by the local training instances with Differential Privacy techniques, to ensure that no data leakage is possible. To our knowledge, this is the first proposal in the field of FL that leverages metadata other than the model’s gradients in order to ensure honesty in the reported local models. Our extensive experiments demonstrate the efficacy of our methods, significantly mitigating the risk of data poisoning while maintaining user privacy and model performance. Our findings suggest that this new user elimination approach serves us with a great balance between privacy and utility, thus contributing to the arsenal of arguments in favor of the safe adoption of FL in safe domains, both in academic setting and in the industry.
	\end{abstract}
}

\begin{document}

\maketitle 


\section{Introduction}

Machine Learning is emerging in becoming a field at the forefront of advancing how we interact with systems and data. It is characterized by the ability of said machines to make data-oriented decisions, while facilitating its users to fulfill decision-making processes, helping them optimize tasks, and enabling a new era of automation.

Due to the rise of available data, researchers and developers are now supplying ML models with large amounts of data, as it is required by a state-of-the art model in order to function and learn correctly. Data typically flow freely and are widely available when it comes to public and everyday tasks, like images, speech, stocks, etc., with people being able to access and use them freely and without any type of license in order to train their own ML models. However, this is not the case for information that is held or produced by a person. Data like facial and private images, health records and location information are and should remain private, as there are not always good intentions by people accessing and analyzing the data. 

The solution to this problem was first proposed in \cite{mcmahan_communication-efficient_2023}, and answers to the name Federated Learning (FL). The goal of FL is to be an efficient and scalable solution to find and collaborate with distributed resources in order to train a Machine Learning model. This is an approach that by its definition allows users to communicate with a central entity and contribute to the learning process while keeping their private data local. This training paradigm, as we will see moving forward, offers potential solutions to the issues haunting traditional machine learning models, such as privacy concerns and high communication costs, while also enabling access to a broader and more diverse range of data sources.

\subsubsection{Problem introduction.}
Nevertheless, there is no panacea in any type of task in computer science, and same goes with Machine Learning and Federated Learning specifically. Letting users actively train a global model seems like a great idea, if of course all of them are totally honest and do not try to act maliciously, or even with curiosity. However, this is not always the case as will be discussed moving forward, as plenty of users may want to either harm or take advantage of the product while participating in the learning process. 

This kind of attacks, namely Data Poisoning Attacks are a broad type of attacks that may contain users trying to alter the labels of their training set, or even the data itself, either with a goal of harming the model, thus they could act in a randomized manner, or in a targeted way in order to manipulate it. During this scenario, we thus take as a given fact that a certain percentage of the users that contribute to the training of the model have malicious intent, thus wanting to poison our system to cause it to wrongly classify instances in the testing phase.

\subsubsection{Motivation and contributions.}

The essential thrust of our research is to contribute in the confrontation of the above-mentioned challenges. In this regard, we are going to examine privacy concerns created by the uncontrolled user participation in FL and present an attack that, as we are going to showcase, threatens the viability of such models. Specifically, we are going to focus on users that try to alter the dataset and thus poison the model, thus launching so called Data Poisoning Attacks. We are going to launch experiments and determine the correct metrics that must be utilized in order to detect such attacks. 

However we are not going to limit our contribution to solely detecting a Data Poisoning Attack in Federated Learning, as we are also going to defend against it and minimize its impact to the final product. Although the field we are describing is rather new and has not yet grown enough, there have been some proposals to detect and prevent Poisoning Attacks. We are going to examine them and then make a valuable contribution to the developing arsenal of defenses proposed by the community, by developing and testing a novel idea which utilizes metadata reported from the users combined with modern Data Privacy techniques, in order for their identity to remain secret. As demonstrated in the paper, the proposed defense mechanisms showcased very positive results for complex image classification tasks, both in model performance and in malicious users' detection.

\section{Preliminaries and Relevant Work}

As we lay the groundwork for our investigation into Poisoning Attacks in Federated Learning, the following section outlines the fundamental concepts and challenges at play. It will provide a comprehensive background on Federated Learning, introduce the critical issue of Poisoning Attacks within this framework, discuss the role of Differential Privacy as a defensive countermeasure, and examine relevant research that has previously tried to tackle the problem.

\subsection{Federated Learning}
Federated Learning is a Machine Learning paradigm where multiple users collaborate to train a model, while each individual's data never leaves their device. The term was first introduced in \cite{mcmahan_communication-efficient_2023} by the Google Research Team, and offered a solution in the problem of decentralized learning, by forcing end-users (e.g. holders of mobile devices) to locally train an instance of the model, update the gradients that were sent to them by a centralized entity and then return their new weights back to it. This approach was conceived due to the significant volume of data on such devices, as well as their substantial computing power. Of course, this computing power is not enough to train a large and scalable ML model, it is however sufficient to train a small dataset with the private data that each device has. 

The training process of an FL model is well described in \cite{liu_distributed_2022}. It begins with a global model being initialized on a central server, with either random or pre-defined weights. This model is communicated by the central server to a subset of the participating devices in the network. Then, each client trains the model that they received locally on their own data, typically over a number of epochs. The training phase is usually similar to traditional machine learning: computing the prediction, comparing it with the true value to compute the loss, and then updating the model parameters using a method like gradient descent to minimize the loss. 

Once local training is complete, each client sends their update, that could include gradients, alterations in weights, or other forms of model parameters, back to the central server. The server then aggregates these updates from all the clients, a process that could be done with a number of methods, with the most simple being FedAvg \cite{mcmahan_communication-efficient_2023}, which consists of an unweighted average of each user's contribution. Finally, the global model is then updated given this aggregated information. Similarly with traditional ML training, the above process is repeated for several rounds until the model's performance reaches a satisfactory level or does not improve significantly. The global model obtained at the end of this process is the final Federated Learning model, which is then subject to evaluation and testing. 

This seems like the optimal solution: personal data never leave users' devices, the server does not need to train locally thus requires less computing power, as that is also distributed to the end users. Most importantly, users are able to use a model that has been trained in a wide variety of data, and not only their own, something that clearly will positively affect the ability of the model to correctly predict new behaviours. 

Federated Learning not only promises the above, but also provides privacy and security both to its end users and their raw data. The avoidance of data flow between server and users is a major step in that direction, but as we will see moving forward, that is not the only one taken. As excellently pointed out in \cite{liu_distributed_2022}, Federated Learning brings the code to the data, instead of bringing the data to the code, something that helps in tackling the problem that legislation like GDPR is trying to prevent. 

\subsection{Differential Privacy}

In this piece of research, we are going to examine, comment and try to tackle problems regarding the protection of user-owned data. In order to do so, it is only fair that we start by providing an introduction to Data Privacy and the most relevant solution to the problem, Differential Privacy. 

One of the fundamental challenges for Privacy Enhancing Techniques has always been the protection of sensitive data. In the era of big-data and personal data collection it is of the utmost importance for companies to ensure their users that their data cannot be directly linked back to them. Moreover, we already have been given an idea of the importance of data privacy in machine learning and in data flow in general. 

Driven by those principles, many approaches have been proposed to the community in an attempt to preserve Data Privacy. The solution was in the making for several years with approaches focusing on the insertion of random noise, most of them from the statistics and databases community, with the most influential being \cite{dinur_revealing_2003}, \cite{dwork_privacy-preserving_2004}, \cite{agrawal_privacy-preserving_2000}. The final and most complete solution came from Dwork in \cite{dwork_algorithmic_2014}, where the principles of a new way of anonymizing data, named Differential Privacy are communicated.

Differential Privacy, as noted from Dwork in her original work, is rather a definition than a strict algorithm. The abstract idea behind Differential Privacy (DP), is that the output of a Differentially Private mechanism, should by independent of whether an individual is present in the domain N. The ”ability” of the adversary to recognize the existence of a column in the dataset, is regulated by the privacy parameter $\epsilon$.  Differential Privacy is defined as following: 
\bigskip
\begin{theorem}{Differential Privacy, given in \cite{dwork_algorithmic_2014}}{}

A randomized algorithm $M$ is $(\epsilon, \delta)$-differentially private, if for all $D_1$ and $D_2$, that differ on at most a single element, and $S\subseteq Range(M)$, stands that:
\begin{center}
    $$ Pr[M(D_1) \in S] \leq e^\epsilon \cdot Pr[M(D_2) \in S] + \delta$$ 
\end{center}
\end{theorem}

The parameter $\epsilon$ can be a regulator to the trade-off between privacy and usability that we mentioned, as lower values of $\epsilon$ mean stronger privacy guarantees and higher values indicate a more usable dataset. The parameter $\delta$ accounts for a small number that is present to even the result when the upper bound does not hold. If $\delta = 0$, we say that $M$ is $\epsilon$-differentially private.  

The most common way that DP is introduced in a dataset or in a learning process, is via introducing random noise to the data. This noise is then “cleared out” via sophisticated aggregation methods which we are going to examine moving forward.  

There are multiple variations of Differential Privacy due to the potential for interpretation. The two main types are Central D.P. and Local D.P. (\cite{cormode_privacy_2018}), which differ primarily in terms of who is responsible for the data. In the Central model, the data curator collects non-private data and applies a D.P. algorithm, requiring a trusted curator. Conversely, in the Local model, the data curator may be untrusted because users apply a specific protocol to perturb their own data.  

In this paper, we are going to focus more on Local DP, because of its alignment with the decentralized tasks that we are dealing with. 

\subsection{Poisoning Attacks}

Federated Learning algorithms are robust for distributed learning, given the hypothesis that participating users are truthful and honest. However, in the sector of Security and in the spectre of this piece of research, this will not be the case, as we will focus on users with malicious intentions.    

Specifically, the threat model for our attacks introduces users whose goal is to harm our models, in a targeted and predefined way as follows: they aim in misleading the model and try to cause it to misclassify a specific class as another \cite{chen_targeted_2017, bagdasaryan_how_2019, fang_local_2021, tolpegin_data_2020}. A valid example could be the case of image recognition, where attackers try to misclassify a specific type of image. If this is implemented in autonomous driving, some users could try to confuse the model by presenting images of bicycles as trunks, thus causing the car systems to malfunction \cite{patel_bait_2020}. 

We can see by the above example that this type of attack is a very serious one, and no one can guarantee that it will not happen, or that it can be controlled in a distributed scenario. This happens because each individual's data is kept local and private, something that allows them to actively lie about their labels, without the centralized authority knowing that fact \cite{bagdasaryan_how_2019}. Intuition could lead us to believe that a significant portion of the users have to collude to have a noticeable impact on the model, but as we will show in later sections, even a small percentage of malicious users can have an impact on the behaviour of the model.  

In the traditional Machine Learning setting it would be easier to detect such attacks as the central entity can access all the data that is used for training, thus catch such anomalies before the training phase. This has been attempted and succeeded in a sufficient level as shown in \cite{goldblum_dataset_2023}. However, Federated Learning is meant to preserve the users' privacy, thus the central authority must not have any kind of access to the dataset that each user utilizes to locally train the model. 

Thus, the question that arises from this setting is: "How can we defend against an attacker that tries to inject faulty data into our model, if we never look at the data?". This is the problem that we will attempt to mitigate throughout this paper. 

In our attempts, we are actually going to add to our arsenal a core setting of FL: data privacy techniques. We are thus going to investigate how we can utilize data privacy in our favor, in order to protect honest users and at the same time detect malicious ones whilst training our Federated Learning Models.

\subsection{Defending against Data Poisoning Attacks}
Before diving into our own research, it is only right that we examine relevant research that is being conducted in the scope of the subset of interest of this paper. The majority of relevant defenses has been applied in the Centralized learning scenario, as the Federated one is rather new, and, as we will see moving forward, more difficult to defend against. The community has explored various methods to tackle the issue, which can broadly be categorized based on whether they involve the elimination of potentially malicious participants. 

Researchers developing algorithms for the first category focus on altering one or more of the layers of the model in order to “inject” the defense directly into the model and disallow attackers from poisoning the model. This can happen with several Privacy Enhancing Techniques, such as Differential Privacy, showcased in \cite{naseri_local_2022, thapa_splitfed_2022, sun_can_2019}, Homomorphic Encryption \cite{chen_federated_2022}, Secure Multiparty Computation \cite{zhang_survey_2021}. 

As shown in the above-mentioned papers, the majority of algorithms succeed in detecting smaller percentages of malicious clients, but most of the times struggle to generalize when more attackers are present. This defense method is less invasive, but could introduce more computational overhead or harm the performance of the primary task, which is the model training.  

The second category is the one that we are going to focus on and includes algorithms that try to detect anomalies in the training phase and eliminate users that create them \cite{tran_spectral_2018, suciu_when_2018, cretu_casting_2008, barreno_security_2010}. This type of defenses requires an extra step in the training process, namely an "anomaly detection algorithm", which is what each solution in that area tries to create. In theory, the more sophisticated the algorithm, the better. However, as seen in relevant literature, an extremely specific algorithm can create the equivalent of overfitting, thus not generalizing well in different and more diverse tasks.

All the above are tested against centralized ML models, while we were unable to find sufficient work in this subset of defenses when it comes to FL systems. A reason for that could be the young age of Federated Learning and relevant attacks in those models. Nevertheless, the main cause of absence of such defenses is the promise of FL for no extra data leakage that can link the user with their data, something that is a problem for implementing all the above papers in that scenario. In this paper, we aim in changing that, by introducing such a defense for FL, while at the same time respecting users’ privacy, by combining both of the above-mentioned techniques: adding an extra layer in our model and implementing an anomaly detection algorithm. 

\section{Poisoning Attacks against Federated Learning}

\subsection{Metrics used}
Throughout this paper we are going to use metrics that will allow us to better comprehend the security and the utility offered by the models that we are going to examine and poison with our experiments. Those metrics are defined above.

\begin{itemize}
    \item \textbf{Sparse Categorical Accuracy}: Used to assess the accuracy of a model's predictions by comparing the predicted class labels to the ground truth labels
    \item \textbf{CrossEntropy Loss}: measures the difference between the predicted probability distribution and the true probability distribution of the classes. In the context of our models, this metric will help us quantify how well the predicted probabilities match the actual class labels
    \item \textbf{Source Class Recall}: The number of correct positive predictions that were made out of all positive predictions that could have been made by the model. When a dishonest user changes the labels of the data, the metric will drop, as fewer (to none) correct positive predictions will be made for the specific class the attacker is trying to poison.
\end{itemize}

\subsection{Experiments results}

In this section, we delve into the experiments of Data Poisoning Attacks against Federated Learning. Our exploration is guided by the above-mentioned metrics, which will provide a comprehensive evaluation of the attack impact and, later on, the effectiveness of our defense strategies. 

We employ two widely recognized datasets, MNIST and CIFAR-10, as the basis for our experiments, leveraging their diverse and complex data. Our analyses and model implementations are conducted using the PyTorch library. The model architectures and hyper-parameter choices are available in the code accompanying this paper, as presented in \cite{galanis_defending_nodate}. 

We are going to train a convolutional neural network (CNN) model on the MNIST dataset for 30 epochs, and for the CIFAR dataset for 60 epochs, parameters that we are also going to use when we apply our defense mechanism. We are going to focus on the impact that the increasing presence of malicious users has in the behaviour and robustness of the model.

\subsubsection{Impact in standard metrics.}

The first metric we used is the Sparse Categorical Accuracy of the model. An initial observation is that the honest model's accuracy follows the normal curve that we are used to seeing in central learning scenarios for both MNIST and CIFAR datasets. In general, the shape of the curves indicates well-trained models, with an appropriate learning rate, sound architecture, correct optimizers and good preprocessing of the data.  

When it comes to the implementation of the attack, a very interesting observation is that the overall accuracy of the models remain at satisfactory levels throughout our experiments. Even with a higher percentage of malicious users, the models still converge to better-than-average validation accuracy  values, as even during the experiments with 50\% of the users being malicious, the metric does not differ more than 10\% from the honest models', something that could be achieved with a model solely with honest users, because of minor issues in architecture or other differences in the ML pipeline. 

\begin{figure}[htb]
    \centering
    \includegraphics[width=0.7\linewidth]{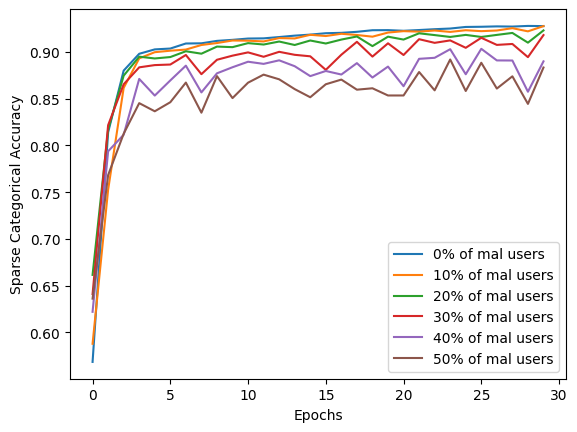}
    \includegraphics[width=0.7\linewidth]{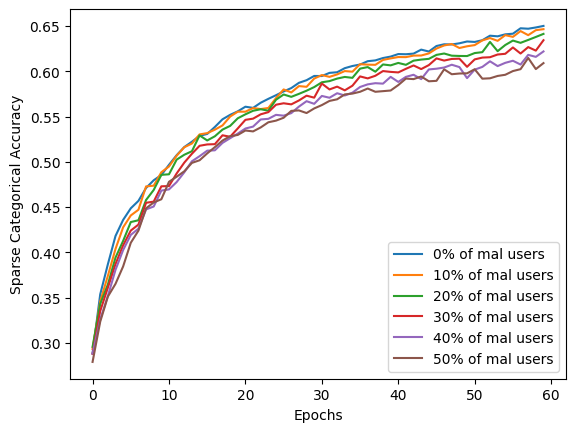}
    \caption{Sparse Categorical Accuracy over the different percentages of malicious users present for MNIST (top) and CIFAR (bottom) datasets.}
    \label{fig:both_images}
\end{figure}

Therefore, it can be inferred that in a real-world scenario where a single model is trained, detecting a poisoned model solely using its testing accuracy becomes exceedingly challenging. To illustrate this, \emph{Figure 2} presents a side-by-side comparison of the accuracy results from two contrasting scenarios in our experiments: a fully honest model (displayed in the top graph) and a model with 50\% malicious client participation (shown in the bottom graph). The comparison reveals that observing differences between these two cases is difficult through visual inspection of the accuracy metric alone, as both curves exhibit similar shapes and converge to comparable accuracy levels after a designated number of epochs.

\begin{figure}[htb]
    \centering
    \includegraphics[width=.8\linewidth]{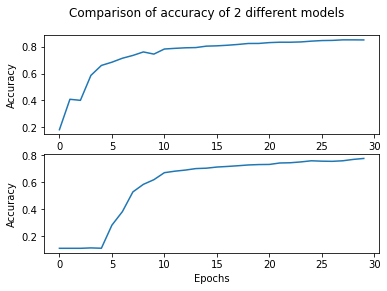}
    \caption{Comparison of the accuracy curve for an honestly and a maliciously trained model}
\end{figure}

The next metric that we are going to comment on is the Crossentropy Loss that we gathered by evaluating our models with the test data. As we can see in \emph{Figure 3} below, again, every model follows the same curve, and with extremely small deviations from the honest model, even with 50\% of the users being adversarial. However, as we will see later on, the ability to distinguish malicious participants (during the training process) due to their slightly higher produced loss, can be an interesting observation. 

\begin{figure}[htb]
    \centering
    \includegraphics[width=0.7\linewidth]{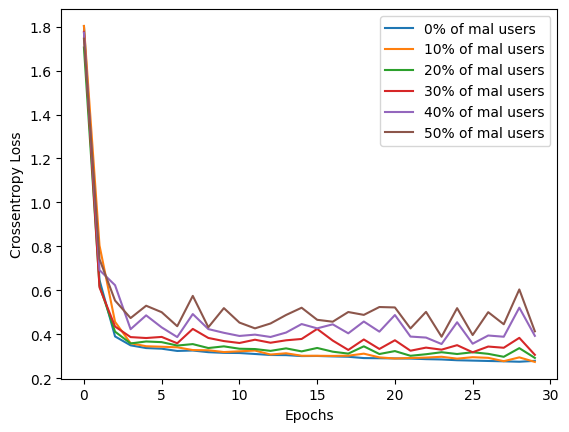}
    \includegraphics[width=0.7\linewidth]{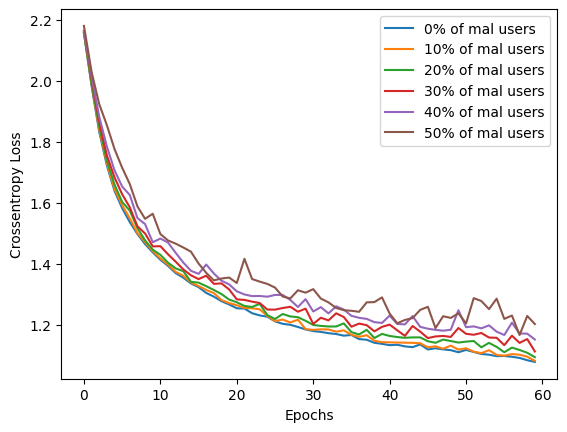}
    \caption{Crossentropy Loss over the different percentages of malicious users present for MNIST (top) and CIFAR (bottom) datasets.}
    \label{fig:both_images}
\end{figure}

\subsubsection{Impact in Source Class Recall.}

The final metric that we gathered while training and evaluating our models was the Recall of the source class, i.e., the class that we attack with a goal to misclassify. It is clear from the graph shown in \emph{Figure 4} that this metric represents our attack's impact accurately.   

\begin{figure}[htb]
    \centering
    \includegraphics[width=0.8\linewidth]{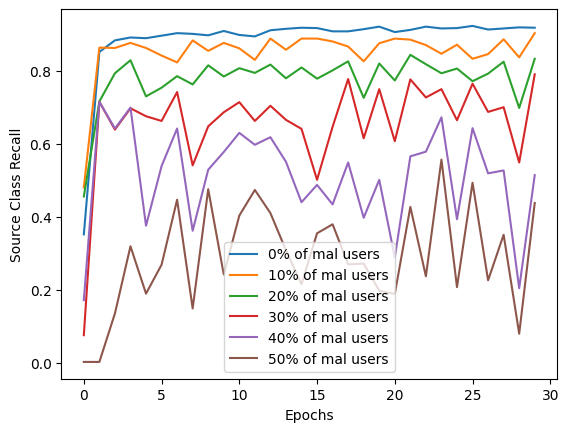}
    \includegraphics[width=0.8\linewidth]{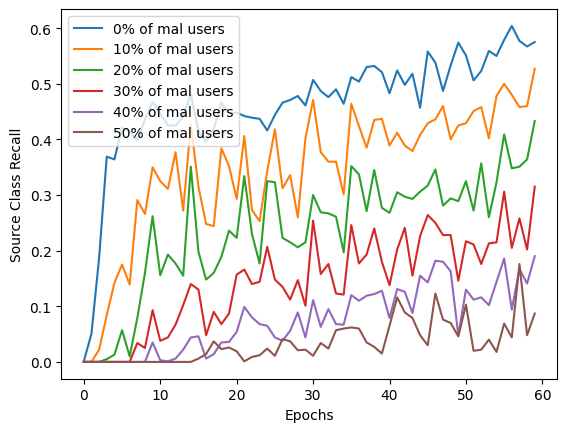}
    \caption{Source Class Recall over the different percentages of malicious users present for MNIST (top) and CIFAR (bottom) datasets.}
    \label{fig:both_images}
\end{figure}

In small percentages of the users being malicious (up to 20\%), we can see that the recall curve is similar to the honest model's one, especially for the MNIST dataset, which is considerably an easier learning task. However, when malicious users become more than 20\% of the total clients, the metric struggles to surpass 0.4 for MNIST and 0.2 for CIFAR, which is an abnormal behaviour for regular training, something that successfully allows us to observe the poisoning attack. For extreme cases (i.e., half of the users being malicious) the metric struggles to get higher than 0.1 for both datasets, something that indicates the total misclassification of the source class, which was the objective of the attack in the first place. 

It is important to highlight that this metric derives from the evaluation of the test set, comprising solely honest labels, for each client. Thus, we are confident to make the observation that if the aggregating authority has access to an honest testing set for the dataset that the model is being trained on, then they could successfully discover a targeted poisoning attack, by computing a single class recall for every class of the task in question. 

Given our above experiments, we can draw the conclusion that it is difficult for one to discover a Targeted Poisoning Attack in a federated scenario. The aggregator only has access to the weights each client returns and thus with the standard federated algorithms available, he must update the central model's state by averaging all the weights that he receives. 

Moreover, the aggregator is not able to detect the attack by utilizing the common metrics that are returned after evaluating the model, as the accuracy and the accumulative loss are not helpful in that direction. A metric that seems to help in that cause is the Recall of the source class of the attack, which produces considerably lower numbers when a high percentage of malicious clients are present. 

Even when an attack is detected, pinpointing the specific users responsible for the poisoning remains infeasible due to the limited identifying data returned by each client. Of course, this is being done due to privacy concerns, as one of the key points of federated is for the clients participating to not be identifiable. It should be noted that, with the current algorithms available, even if the aggregator could identify the malicious clients, it would only be after their weights have been integrated into the global model. Consequently, this allows only for detection of the attack post-facto, rather than its prevention.

The objective of this paper is to address this challenge by developing methods to detect and defend against such attacks proactively, thereby preventing their impact on the global model. In the following section, we will present and discuss the defense mechanisms we have devised and implemented to achieve this goal.

\section{Novel Algorithm for Defending against Poisoning Attacks}

In previous chapters we showcased the severity and impact that a targeted poisoning attack can have on a Federated Learning model. In this one, we are going to find a way to tackle it, with the end-goal of eliminating the users that try to poison our model.  

Our approach to defending against these attacks adopts an innovative perspective, deviating from methodologies observed in the literature. During the literature review we observed the pattern of defense mechanisms that have been adopted by researchers in the field, which does not include the user reporting anything else than the gradients back to the aggregator. This is done due to privacy concerns of the user being identified by any other metadata that they may report. 

However, we opted to make the users return their training loss for their local training round. We assume that the training loss for users that act maliciously will behave differently than the honest ones, thus by aggregating this information we will be able to detect them and eliminate them from contributing to the training phase of the model. In later sections we will confirm that allegation by observing the behaviour of the model when this piece of information is utilized. To the best of our knowledge, as of February 2024, there is no published work utilizing loss metrics to distinguish between malicious and honest clients.

It is clear that if a user reports the exact value of their training loss, this could prove catastrophic, as somebody could extract useful information regarding the instances that the user used for training, something that breaks the promise of the privacy offered to the users. In order to avoid that, we are going to utilize the foundations and the logic behind Local Differential Privacy, by injecting a random amount of noise every time a user reports their loss. 

\subsection{Threat Model}
To formally describe the algorithm and the logistics of our defense solution, we must first describe the threat model under which we are operating.  

As we have already established, the attack scenario occurs in a Federated Learning context, where users have the responsibility of training a local model which is then communicated to a central authority in charge of aggregating an upgrading the global model with the gradients given by the users. Thus, the users are totally independent and decentralized, something that leads to the server not having any information about their training, other than the values reported by them. In our case, this information includes the weights shaped by the training, and the CrossEntropy Loss of the local model as a result of training the user's dataset.  

We assume that the user reports a correct value for both above-mentioned elements, as this is crucial for our defense algorithm to function correctly. This can be easily ensured in a real world scenario, by the correct development of the framework, or by introducing cryptographic primitives that help in that context, such as Zero Knowledge Proofs or Commitment Schemes \cite{schnorr_efficient_1990}, which of course add computational overhead, but at the same time ensure that a malicious user will not succeed in reporting a false value. This paper does not delve into the industrial implementation of the solution; therefore, the focus is not on the practical aspects of ensuring the reliability of user-reported data.

We also assume that the server is not actively malicious and not colliding with malicious users, as an arbitrary acting aggregator could ignore the algorithm of the defense and only include malicious individuals in the global training step.

When it comes to percentages of the participating users being actively malicious, there is no limit, as we are going to examine numbers ranging from 0\% up to high percentages. However, as we have already seen, there is no point in raising the percentage higher than 40\%, as it makes no difference to the already harmed model. Hence, we are going to assume the maximum percentage of malicious users participating in an FL training process as 40\%, and point out that for higher numbers than those, the defense algorithm will work but will have worse results. 

Finally, the definition of a "malicious user" expands as a device participating in the training procedure that is totally controlled by an adversary, who can view, alter labels of already existing instances, as well as insert new instances with new, false labels. This could be accomplished either by physical or remote access of the attacker to the victim's device. 

\subsection{Hyper-parameters used}

To ensure transparency and provide clear insights into the methodology of our experiments, below we present \emph{Table 1}, detailing the hyperparameters we employed. This matrix is designed to explain the choices made in tuning the model for both launching and defending against Poisoning Attacks in FL, for both of the datasets we are going to train our models against: MNIST and CIFAR-10.

\begin{table}[h]
\centering
{\footnotesize 
\begin{tabular}{|l|c|c|}
\hline
\textbf{Parameter}             & \textbf{MNIST} & \textbf{CIFAR-10} \\ \hline
Global training epochs      & 30             & 60             \\ \hline
Number of training clients       & 50             & 100            \\ \hline
Number of total clients        & 500            & 500            \\ \hline
Client learning rate           & 0.01           & 0.001          \\ \hline
Base Federated algorithm       & FedAvg       & FedAvg         \\ \hline
Client training epochs         & 10             & 20             \\ \hline
\end{tabular}
}
\caption{Hyperparameters used in training and defending the Federated Learning model against Poisoning Attacks.}
\label{table:hyperparameters}
\end{table}

When it comes to client selection, this is done randomly, based on the Gaussian distribution, which results in both clients that are selected to train in each round, and attackers selected to be random. However, during our experiments, the attackers are a fixed set of users, that does not change through the epochs, in order to better mimic the behaviour of real users. Each client holds a random, equally distributed subset of the training set.

We should also note that due to the randomness introduced by the user selection and the Differential Privacy algorithms, the experiments were run multiple times (10 for each dataset), in order to cancel out any noise or extreme values that could be introduced by that uncertainty in the generated noise.

\subsection{Novel Federated Learning Algorithm}

From the introduction given, it is clear that some alterations to the FL training algorithm must be made for our defense idea to be implemented. In this section we are going to state in detail the way those alterations will result in a new FL algorithm. 

\subsubsection{Local Training Step.}  

The Local Training step involves individual users training a distributed model with their data, following specific protocols to ensure data privacy and model integrity. In this process, being carried out by all the users randomly selected to participate in a round of global training of the Federated Learning Model, the following steps are being carried out:   

\begin{itemize}
    \item The client receives the local model from the centralized entity in charge of coordinating the FL procedure. 
    \item The user relies on the hyperparameters decided and trains the local model with their data. 
    \item During this process, the training loss is monitored, reflecting the updates made to the local model's gradients.
    \item  After completing the training process, the user locally adds to the training loss gathered a quantity of random noise generated by an already known distribution, with predefined bounds, that follows the foundations of Local Differential Privacy.
    \item  Finally, the user reports back to the server the gradients forming the updated version of the local model, as well as the loss value after the insertion of random noise, and nothing else that will help the centralized authority in recognizing or gathering extra information about the user. 
\end{itemize}

\subsubsection{Global Training Step.}  

In the Global Training step, the central server aggregates inputs from various users, applies a decision-making algorithm to identify and exclude potentially malicious contributions, and updates the global model accordingly. In our novel version of the Global Training Step, the following process is carried out: 

\begin{itemize}
    \item A random portion of the total users participating in the training procedure is selected for training in the specific round. 
    \item The global model from the last epoch of global training (or the initialized one if we are training for the first time) is sent to the selected users, where the above-mentioned local training algorithm is enforced. 
    \item The server receives as a tuple the updates from each one of the users participating in the specific round. The tuple includes the weights reported back and the training loss reported by the user. 
    \item The server gathers the losses in a data structure (as simple as a list), while keeping track of the correlation of each loss with the weights reported. 
    \item A specific elimination algorithm is utilized in order for the server to decide on the users that are going to be banned from the update process. 
    \item The clients whose loss do not meet the criteria set by that algorithm are eliminated from the global update of the model, and their identifiers are given to the aggregator in order to be excluded. 
    \item The aggregator given the (predicted by the algorithm) honest users aggregates the global model by utilizing a previously decided algorithm, in the same manner as regular federated learning     
\end{itemize}

In the course of this research, multiple algorithms were evaluated to identify the most effective method for eliminating potentially malicious users from the Federated Learning process. After extensive experimentation and analysis, one of the following algorithms emerged as significantly more successful than others. In this section we will focus on analyzing all the algorithms, as well as presenting the most successful one's results and extracting conclusions based on them. The experimental results of other attempts are available in \emph{Appendix A}.

\subsection{Defense Algorithms}
This section delves into various defense algorithms that will be evaluated for their effectiveness in defending against Data Poisoning Attacks in the Federated scenario.

\subsubsection{Threshold-based eliminating.}

The first, and most simple function consists of eliminating a certain percentage of the users. The clients are sorted based on the reported losses, and the last n\% of them is being eliminated from the global training process. This approach is based on the premise that malicious users are likely to induce higher training losses, thus falling into the lower-performing segment of participants.

\subsubsection{Distance-based eliminating.}

The next function is independent from a fixed percentage, and its goal is to detect the turning point in the sorted list of losses where the clients become malicious. This would function ideally if all the honest users reported significantly less loss than malicious ones, which, even based on our past experiments, cannot be guaranteed to occur consistently. However, this will be a point of observation made clear by using the percentage of correctly spotted attackers later, during our experiments.

\subsubsection{Statistical-based eliminating: Z-Score.}
The next function that we will consider is based on statistical observations, as it takes into account the distribution of the losses from each client. The Z-Score \cite{witte_statistics_2017}, has its roots in the theory of probability and statistics. It provides a measure of how far a given data point deviates from the mean, in terms of standard deviations. Mathematically, the Z-Score $z$ for a data point $x$ is computed as: 

\begin{align*}
z = \frac{x - \mu}{\sigma}    
\end{align*} 

where $\mu$ is the mean of the data and $\sigma$ is the standard deviation. 

The underlying assumption of the Z-Score method is the Central Limit Theorem, as stated in \cite{witte_statistics_2017}, which posits that the sum of a large number of independent and identically distributed variables will be approximately normally distributed, regardless of the original distribution of the variables. Thus, in scenarios where the majority of the data (in our case, the reported training losses) follows a normal distribution, data points that significantly deviate from the mean become statistically notable. 

For our purposes, if the absolute Z-Score of a client's training loss exceeds a predefined threshold, which we are going to set to 1 for 68\% confidence, the client is flagged as an outlier. This criterion is based on the empirical rule which states that for a normal distribution, about 68\% of the data falls within one standard deviations from the mean. 

In applying the Z-Score method to the context of our research, we hypothesize that training losses deviating significantly from the mean are indicative of malicious behavior.

\subsubsection{Clustering-based eliminating: K-Means.}
K-means clustering \cite{hartigan_algorithm_1979} is a type of unsupervised ML algorithm that partitions a dataset into $K$ distinct, non-overlapping clusters. The goal of the algorithm is to minimize the variance within each cluster and maximize the variance between the clusters. To achieve that, it defines clusters such that the total intra-cluster variation, or the sum of squared distances (based on the Euclidean Distance) from the mean of the cluster, is minimized. Despite its simplicity, the K-means algorithm can be very effective and robust, especially when the structure of the data is well-defined and can be roughly easily distinctive. 

In the context of our solution, we aim to leverage the K-means technique to distinguish between honest and malicious clients based on their training losses, by defining two clusters in which the two types of clients will fall, with an end-goal of categorizing them correctly. 

Given the scope of this paper, a detailed exploration of the specific algorithms and mechanics of K-means clustering is beyond our purview. Readers who are interested may refer to \cite{hartigan_algorithm_1979} for an in-depth analysis.

\subsection{Insertion of Local Differential Privacy}

We already mentioned that giving a centralized entity direct access to metadata produced by users while training is breaking the promise of Federated Learning regarding the privacy of the participants. Given the exact value of the training loss, an adversary could make discoveries regarding the data distribution and the data points that each user holds and thus deanonymize the user. However, a small alteration of this reported loss value could solve the problem, as the user will no longer be identifiable by it. 

As we have seen in previous chapters, a simple yet robust tool to protect individual data is Local Differential Privacy (LDP), a solution that can be adopted by each user in order to anonymize their participation. Specifically, LDP pertains to introducing randomness at the individual data level before any aggregation or computation by a central authority takes place. In this setting, each individual's data is perturbed in a way that provides a certain privacy guarantee, represented by an epsilon value. The smaller the $\epsilon$, the stronger the privacy guarantee, as we have already seen in our introduction.

In our case, we opt to use the Laplace mechanism for our task, in order to add noise from the Laplace distribution to the data. The scale of the injected noise is determined by the desired $\epsilon$ value and the sensitivity of the function. 

\subsubsection{Defining the Sensitivity.}

The sensitivity, $\Delta f$, is the maximum amount the output can change by altering a single item in the dataset. Given our data, which are float values representing losses and not distinctive values, the sensitivity should be redefined based on the context.  

In our distributions, the data represents measurements which we are trying to ensure that their perturbation will not drastically change their interpretation by the central entity, and thus the sensitivity is deemed as the smallest change that we want the server to be able to distinguish. Given our previous knowledge from conducting experiments, along with the standard ML pipeline and the reported losses for the datasets we will utilize, we can safely say that even a change of 0.001 can alter the data in a severe way. Therefore, we have determined the sensitivity to be $\Delta f = 0.0001$ based on our data and experimental insights.

\subsubsection{Defining the epsilon value.}

The final parameter we must decide on is the epsilon value, which is responsible for balancing the utility and privacy offered to our users. The largest the epsilon value, the lower the privacy guarantees, but at the same time, the higher the utility offered. A fair balance that is commonly used is to set $\epsilon = 1$. However, we must note that this should not be absolute, and each one creating an LDP algorithm for a similar task could alter it depending on their needs. 

\subsubsection{Defining the LDP algorithm.}

Thus, given a function $f$ with sensitivity $\Delta f$, the scale $b$ for the Laplace distribution is defined as: 

\begin{align*}
b = \frac{\Delta f}{\epsilon}    
\end{align*}

\section{Experimental Results}

In this section, we present the results from experiments conducted while applying our defense solution during the training of a Federated Learning model, in scenarios where different percentages of malicious users are present and perform a targeted Data Poisoning Attack.

In our exploration, the four above-mentioned different defense algorithms were tested for their efficacy against targeted Poisoning Attacks in a Federated Learning context. While all four algorithms provided valuable insights, this section will focus primarily on presenting the results of the most effective algorithm, namely the approach utilizing K-Means. Detailed results and analyses of the other three algorithms have been included in the appendix for reference. This approach allows us to highlight the most impactful findings while making comprehensive data available for further review.

\subsection{Metrics used}

Before conducting the experiments, we must define, as we did in previous chapters, the metrics that are going to be utilized to extract information about the attack and the effect of the defense. Those metrics are: 
\subsubsection{Already-used metrics.}
The classic indicators that we have already seen while launching the attacks, namely the \emph{Sparse Categorical Accuracy}, the \emph{CrossEntropy Loss} and the \emph{Recall of the source class} of the attack. Given our findings in previous sections, which indicated that the first two were not so useful in detecting poisoning attacks, we are mainly going to focus on the last one and compare the Recall of the source class when the defense is present and absent.
    
\subsubsection{Attacker-centric metrics.} In our solution we strive to strike an optimal balance between security and utility during the model training process. This leads us to rating our solution based on metrics that are focused on the attackers' detection, namely the \emph{Accuracy} and the \emph{F1 score} as applied to to malicious users' prediction by the clustering algorithm. As a result, we are going to present the mean value that those two metrics will have during training for each different experiment. 

Those measurements are essential for our work, as their low values would indicate a compromise either in the security or in the utility aspect of our experiments. For example, if the algorithm eliminated 90\% of the users, this could be beneficial for the global model, as no attackers would be present at all, but at the same time it would severely harm its utility, as honest users would be eliminated from the process. Thus, those two indicators will help us confirm that our proposal does not harm the model's training phase, while also preserving the integrity of the users used in this step. 

\subsection{Results on model performance}

The first plots that we observe in \emph{Figures 5 and 6}, demonstrate the metrics that we have used throughout this paper in order to compare the success of poisoning. 
\begin{figure}[htb]
    \centering
    \includegraphics[width=0.8\linewidth]{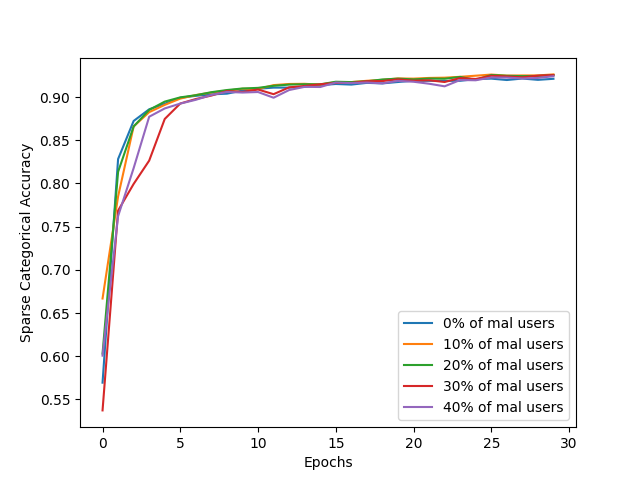}
    \includegraphics[width=0.7\linewidth]{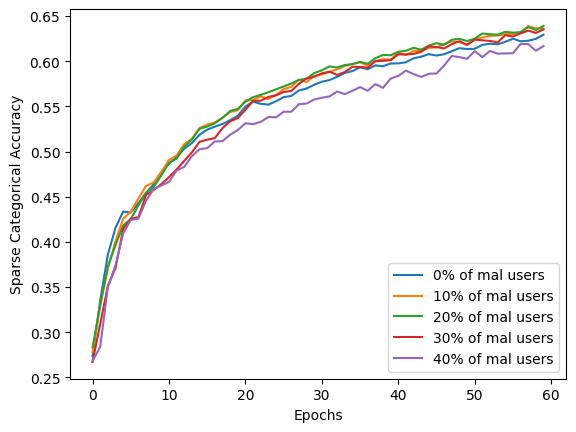}
    \caption{Accuracy over the different percentages of malicious users present for MNIST (top) and CIFAR (bottom) datasets.}
    \label{fig:both_images}
\end{figure}

\begin{figure}[htb]
    \centering
    \includegraphics[width=0.9\linewidth]{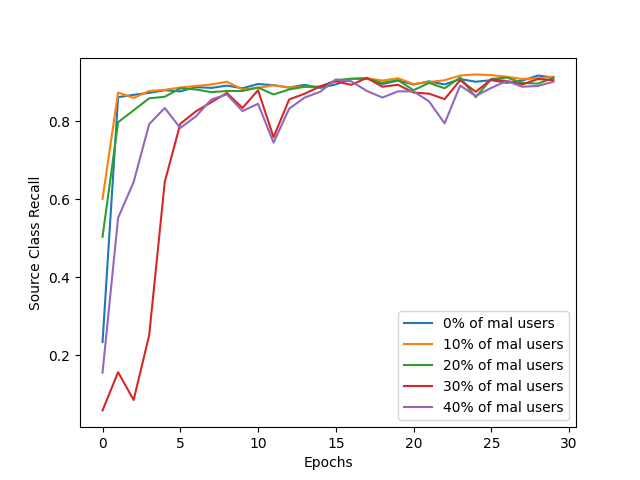}
    \includegraphics[width=0.8\linewidth]{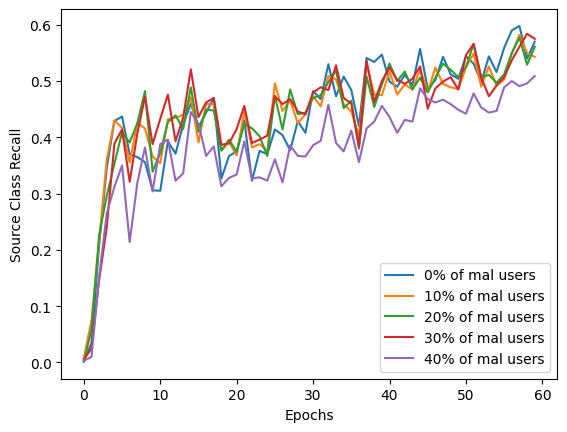}
    \caption{Source Class Recall over the different percentages of malicious users present for MNIST (top) and CIFAR (bottom) datasets.}
    \label{fig:both_images}
\end{figure}

\begin{figure}[htb]
    \centering
    \includegraphics[width=0.85\linewidth]{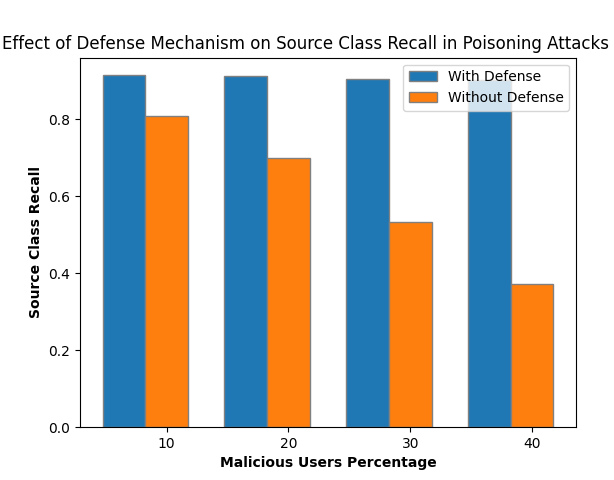}
    \includegraphics[width=0.85\linewidth]{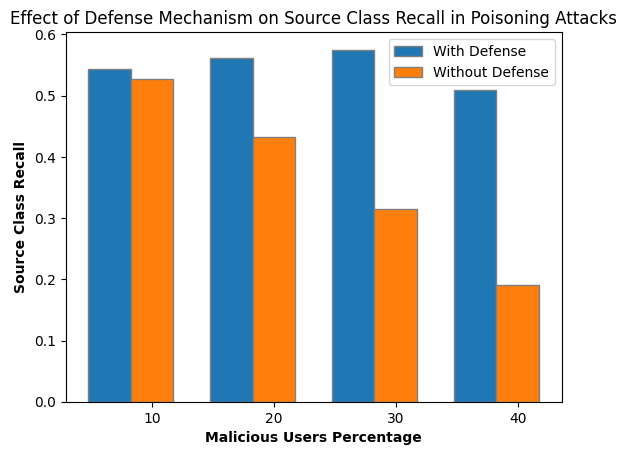}
    \caption{Comparing Source Class Recall with and without the defense mechanism for MNIST (top) and CIFAR (bottom) datasets.}
    \label{fig:both_images}
\end{figure}

Remarkably, even with 40\% of users being malicious, the implementation of our elimination-oriented defense results in negligible deviation from the performance of the original, honest model. A small difference was always present in the accuracy function, but the most encouraging behaviour is observed in the Source Class Recall metric, where all five models behave similarly, from the beginning of training up until the last epoch. This is a strong indication that the K-means algorithm is able to detect the vast majority of malicious users starting even from the $1^{st}$ epoch. Of course, in the extreme case of 40\% of the clients being malicious, despite the similar shape of the Recall curves, we observe a slight decrease in the values of the metric, especially in the CIFAR dataset, which is a rather more complex dataset when it comes to learning.

However, as seen in \emph{Figure 7} below, which presents the values of the Source Class Recall when our defense is present and absent, we can clearly state that even for 40\% of the clients being malicious there is a significant improvement in the metric, as when there is no defense applied, it struggles to surpass very low standards for both of our datasets, which is definitely not the case for when we enforce our defense mechanism.

These results indicate that the models we trained under our defense algorithm perform similarly to their honestly trained counterparts for each dataset. This leads us to believe that during the training phase, our defense mechanism effectively identified and eliminated users exhibiting anomalously high differentially private training loss. Consequently, this maintained the model in a sufficiently honest state, allowing it to yield the above-shown metrics. From the perspective of model utility, this is a highly favorable outcome, as it demonstrates our algorithm's ability to safely train a Federated Learning model in environments with potential attackers, ensuring both efficiency and robustness in the model's test performance.

\subsection{Results on attacker detection}

However, our exploration does not end here. The next crucial aspect of our research involves investigating the attacker-specific metrics. Our objective is to confirm that our algorithm not only accurately identifies malicious participants but also maintains a satisfactory F1 score for that task. This focus ensures that while malicious users are reliably eliminated, the contribution of honest users is preserved, optimizing the utility of their data in the training process. The above metrics are drawn as the mean value throughout the range of epochs of training the model, and thus clustering the users as malicious or honest.

\begin{figure}[htb]
    \centering
    \includegraphics[width=0.7\linewidth]{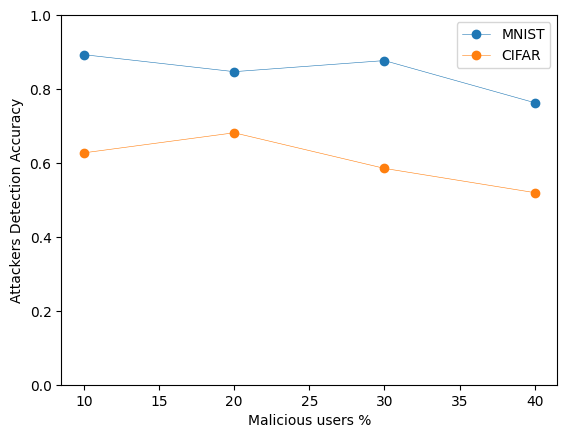}
    \includegraphics[width=0.7\linewidth]{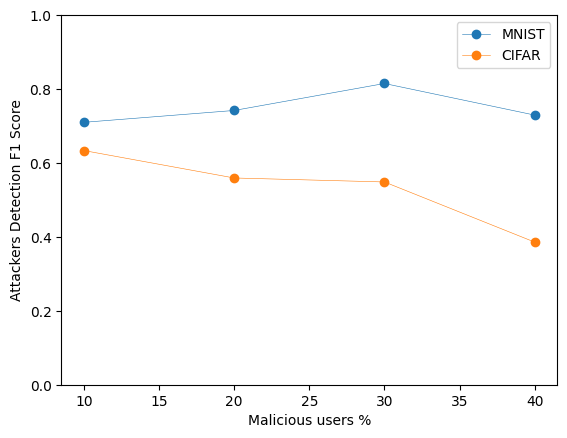}
    \caption{Accuracy (top) and F1 Score (bottom) for the task of classifying malicious users as such during training.}
    \label{fig:both_images}
\end{figure}

As we can see in \emph{Figure 8}, the detection algorithm does a good job in detecting the presence of malicious clients both in MNIST and in CIFAR datasets.

The top graph presents the attacker detection accuracy for two datasets, against the percentage of malicious users present. For the MNIST dataset, the detection accuracy remains relatively high and consistent, only showing a marginal decrease as the proportion of malicious users increases to 40\%. For the CIFAR dataset we observe a more pronounced decline as the percentage of malicious users grows more than 30\%, suggesting that the complexity of the dataset may insert additional challenges in accurately identifying malicious users' presence. 

The bottom graph showcases the F1 score of the task of detecting malicious users, again in comparison with the percentage of the actual percentage of attackers in our model. This metric further underscores the performance of our defense strategy, due to its identity, namely balancing the precision and recall of our detection algorithm. As illustrated in the graph, the MNIST dataset maintains a robust F1 score across all levels of attacker presence, with only a slight decrease even as the percentage of malicious users reaches 40\%. At the same time, the CIFAR dataset's F1 score reveals a larger decline for the maximum value of malicious users percentage in our experiments, indicating a reduction in the balance between precision and recall. 

The common ground for both datasets is that they behave satisfyingly and similarly for malicious users presence up to 30\%, while declining after that, which is anticipated, as we have previously observed during the model performance experiments. However, when almost half of the users are malicious, the prediction continues to produce a satisfying accuracy for each of the datasets (over 55\%), something that confirms the success of our algorithm to detect such users even under extreme circumstances in difficult learning scenarios, such as the CIFAR dataset.

Finally, an important observation while considering and comparing both experiments, is the following: despite the lower accuracy and F1 score produced for the task of attacker detection in the case of 40\% malicious user participation in the difficult task of predicting the CIFAR dataset, the model behaviour when the defense is present (our primary task), is still of high quality. We can observe that from the fact that the Source Class Recall produced is almost equal to the level of the experiment with 10\% malicious user presence.

\section{Discussion and Conclusion}

Our research was focused on exploring the emerging field of Federated Learning and a possible set of attacks against it, namely Data Poisoning Attacks. We began by launching several such attacks against FL models and observing their behaviour and their impact in the model through detailed experiments, while defining key metrics that would help us tackle them. We found out that without harming the model's accuracy, they can actually alter the predictions, especially if the volume of malicious users is over 20\%. 

Upon discovering the subtle yet significant impact of malicious users, our research pivoted towards deploying a defense mechanism that preserves the foundations of FL: accuracy and privacy. We proceeded in observing that although the overall loss reported is not altered, the users with malicious intentions tend to report a higher loss value than the honest ones. 

The breakthrough of this paper is that we took advantage of this reported loss by each user in order to predict the malicious ones and eliminate them from the training procedure. We did that while being able to respect the participants' privacy, one of the core promises of Federated Learning.

We found an elegant and robust way to do so, by introducing an extra layer of Local Differential Privacy based on the LaPlace distribution before the users reported their loss back to the server. We then combined the idea of predicting and eliminating malicious users with the privacy layer and came up with a clustering-based algorithm in order to classify users as attackers. We presented experiments that showcased how well we succeeded in this task, in two well-know datasets in the field of Machine Learning, namely MNIST and CIFAR. We conducted several experiments to prove the success of our solution, both regarding model behaviour, and correct user elimination. By doing so, we are positive that this method is a valid and robust defense against label Data Poisoning FL models.

Since we introduced a new way for defending against Poisoning Attacks, there is definitely space for future work that could be conducted. The goal of this scientific work was not to provide an exhaustive way of solutions for classifying users as malicious or honest, but rather to introduce the innovative above-mentioned method of eliminating users based on their loss. We did actually provide a successful way of carrying out the classification task with the K-Means algorithm, but there are many directions that could be further explored.

Finally, given that our defense is not specific or tight to any model dataset or learning scenario, this mechanism could be used to defend against different type of attacks in alternate architectures, with the prerequisite that the attacks increase the reported loss during training.

\subsection{Acknowledgments}

This work would not have been possible if not for some people offering their insights and knowledge of the field. I would like to thank Professor Emiliano De Cristofaro of University of California, Riverside, my MSc Thesis supervisor at UCL, for introducing me to the world of Federated Learning, and for pushing me to my limits when it comes to innovation and ideas for this extremely interesting subject. Moreover, I would like to thank Professor Lorenzo Cavallaro of UCL, who closely monitored my research and provided me with ideas and guidance throughout my experiments and the publication of this paper. I would also like to thank all the people in the Federated Learning and in the Privacy community, with whom I have had great discussions over the past two years. Last but not least, I would like to thank my current company, Canonical, for allowing me to pursue my research interests, and supporting me throughout this journey, while trying to make the world more open-source and more secure.


\printbibliography

@misc{galanis_defending_nodate,
	title = {Defending agaist {Poisoning} {Attacks} in {FL} (code repository)},
	url = {https://github.com/nikosgalanis/data-poisoning-defense-fl},
	language = {en},
	urldate = {2024-04-14},
	journal = {GitHub},
	author = {Galanis, Nick},
}

@misc{tran_spectral_2018,
	title = {Spectral {Signatures} in {Backdoor} {Attacks}},
	url = {http://arxiv.org/abs/1811.00636},
	doi = {10.48550/arXiv.1811.00636},
	urldate = {2023-09-01},
	publisher = {arXiv},
	author = {Tran, Brandon and Li, Jerry and Madry, Aleksander},
	month = nov,
	year = {2018},
	note = {arXiv:1811.00636 [cs, stat]},
}

@inproceedings{suciu_when_2018,
	title = {When {Does} {Machine} {Learning} \{{FAIL}\}? {Generalized} {Transferability} for {Evasion} and {Poisoning} {Attacks}},
	isbn = {978-1-939133-04-5},
	shorttitle = {When {Does} {Machine} {Learning} \{{FAIL}\}?},
	url = {https://www.usenix.org/conference/usenixsecurity18/presentation/suciu},
	language = {en},
	urldate = {2023-09-01},
	author = {Suciu, Octavian and Marginean, Radu and Kaya, Yigitcan and Iii, Hal Daume and Dumitras, Tudor},
	year = {2018},
	pages = {1299--1316},
}

@inproceedings{cretu_casting_2008,
	title = {Casting out {Demons}: {Sanitizing} {Training} {Data} for {Anomaly} {Sensors}},
	shorttitle = {Casting out {Demons}},
	doi = {10.1109/SP.2008.11},
	booktitle = {2008 {IEEE} {Symposium} on {Security} and {Privacy} (sp 2008)},
	author = {Cretu, Gabriela F. and Stavrou, Angelos and Locasto, Michael E. and Stolfo, Salvatore J. and Keromytis, Angelos D.},
	month = may,
	year = {2008},
	note = {ISSN: 2375-1207},
	pages = {81--95},
}

@article{barreno_security_2010,
	title = {The security of machine learning},
	volume = {81},
	issn = {0885-6125, 1573-0565},
	url = {http://link.springer.com/10.1007/s10994-010-5188-5},
	doi = {10.1007/s10994-010-5188-5},
	language = {en},
	number = {2},
	urldate = {2023-09-01},
	journal = {Machine Learning},
	author = {Barreno, Marco and Nelson, Blaine and Joseph, Anthony D. and Tygar, J. D.},
	month = nov,
	year = {2010},
	pages = {121--148},
}

@misc{naseri_local_2022,
	title = {Local and {Central} {Differential} {Privacy} for {Robustness} and {Privacy} in {Federated} {Learning}},
	url = {http://arxiv.org/abs/2009.03561},
	urldate = {2023-09-01},
	publisher = {arXiv},
	author = {Naseri, Mohammad and Hayes, Jamie and De Cristofaro, Emiliano},
	month = may,
	year = {2022},
	note = {arXiv:2009.03561 [cs]},
}

@article{goldblum_dataset_2023,
	title = {Dataset {Security} for {Machine} {Learning}: {Data} {Poisoning}, {Backdoor} {Attacks}, and {Defenses}},
	volume = {45},
	issn = {1939-3539},
	shorttitle = {Dataset {Security} for {Machine} {Learning}},
	doi = {10.1109/TPAMI.2022.3162397},
	number = {2},
	journal = {IEEE Transactions on Pattern Analysis and Machine Intelligence},
	author = {Goldblum, Micah and Tsipras, Dimitris and Xie, Chulin and Chen, Xinyun and Schwarzschild, Avi and Song, Dawn and Mądry, Aleksander and Li, Bo and Goldstein, Tom},
	month = feb,
	year = {2023},
	note = {Conference Name: IEEE Transactions on Pattern Analysis and Machine Intelligence},
	pages = {1563--1580},
}

@misc{patel_bait_2020,
	title = {Bait and {Switch}: {Online} {Training} {Data} {Poisoning} of {Autonomous} {Driving} {Systems}},
	shorttitle = {Bait and {Switch}},
	url = {http://arxiv.org/abs/2011.04065},
	urldate = {2023-09-01},
	publisher = {arXiv},
	author = {Patel, Naman and Krishnamurthy, Prashanth and Garg, Siddharth and Khorrami, Farshad},
	month = dec,
	year = {2020},
	note = {arXiv:2011.04065 [cs]},
}

@misc{fang_local_2021,
	title = {Local {Model} {Poisoning} {Attacks} to {Byzantine}-{Robust} {Federated} {Learning}},
	url = {http://arxiv.org/abs/1911.11815},
	doi = {10.48550/arXiv.1911.11815},
	urldate = {2023-09-01},
	publisher = {arXiv},
	author = {Fang, Minghong and Cao, Xiaoyu and Jia, Jinyuan and Gong, Neil Zhenqiang},
	month = nov,
	year = {2021},
	note = {arXiv:1911.11815 [cs]},
}

@article{hartigan_algorithm_1979,
	title = {Algorithm {AS} 136: {A} {K}-{Means} {Clustering} {Algorithm}},
	volume = {28},
	issn = {00359254},
	shorttitle = {Algorithm {AS} 136},
	url = {https://www.jstor.org/stable/10.2307/2346830?origin=crossref},
	doi = {10.2307/2346830},
	language = {en},
	number = {1},
	urldate = {2023-08-25},
	journal = {Applied Statistics},
	author = {Hartigan, J. A. and Wong, M. A.},
	year = {1979},
	pages = {100},
}

@book{witte_statistics_2017,
	title = {Statistics},
	isbn = {978-1-119-25451-5},
	language = {en},
	publisher = {John Wiley \& Sons},
	author = {Witte, Robert S. and Witte, John S.},
	month = jan,
	year = {2017},
	note = {Google-Books-ID: KcxjDwAAQBAJ},
}

@inproceedings{schnorr_efficient_1990,
	address = {New York, NY},
	series = {Lecture {Notes} in {Computer} {Science}},
	title = {Efficient {Identification} and {Signatures} for {Smart} {Cards}},
	isbn = {978-0-387-34805-6},
	doi = {10.1007/0-387-34805-0_22},
	language = {en},
	booktitle = {Advances in {Cryptology} — {CRYPTO}’ 89 {Proceedings}},
	publisher = {Springer},
	author = {Schnorr, C. P.},
	editor = {Brassard, Gilles},
	year = {1990},
	pages = {239--252},
}

@inproceedings{agrawal_privacy-preserving_2000,
	address = {New York, NY, USA},
	series = {{SIGMOD} '00},
	title = {Privacy-preserving data mining},
	isbn = {978-1-58113-217-5},
	url = {https://dl.acm.org/doi/10.1145/342009.335438},
	doi = {10.1145/342009.335438},
	urldate = {2023-03-16},
	booktitle = {Proceedings of the 2000 {ACM} {SIGMOD} international conference on {Management} of data},
	publisher = {Association for Computing Machinery},
	author = {Agrawal, Rakesh and Srikant, Ramakrishnan},
	month = may,
	year = {2000},
	pages = {439--450},
}

@inproceedings{cormode_privacy_2018,
	address = {New York, NY, USA},
	series = {{SIGMOD} '18},
	title = {Privacy at {Scale}: {Local} {Differential} {Privacy} in {Practice}},
	isbn = {978-1-4503-4703-7},
	shorttitle = {Privacy at {Scale}},
	url = {https://dl.acm.org/doi/10.1145/3183713.3197390},
	doi = {10.1145/3183713.3197390},
	urldate = {2023-03-24},
	booktitle = {Proceedings of the 2018 {International} {Conference} on {Management} of {Data}},
	publisher = {Association for Computing Machinery},
	author = {Cormode, Graham and Jha, Somesh and Kulkarni, Tejas and Li, Ninghui and Srivastava, Divesh and Wang, Tianhao},
	month = may,
	year = {2018},
	pages = {1655--1658},
}

@article{dwork_algorithmic_2014,
	title = {The {Algorithmic} {Foundations} of {Differential} {Privacy}},
	volume = {9},
	issn = {1551-305X},
	url = {https://doi.org/10.1561/0400000042},
	doi = {10.1561/0400000042},
	number = {3–4},
	urldate = {2023-03-16},
	journal = {Foundations and Trends® in Theoretical Computer Science},
	author = {Dwork, Cynthia and Roth, Aaron},
	month = aug,
	year = {2014},
	pages = {211--407},
}

@inproceedings{dwork_privacy-preserving_2004,
	address = {Berlin, Heidelberg},
	series = {Lecture {Notes} in {Computer} {Science}},
	title = {Privacy-{Preserving} {Datamining} on {Vertically} {Partitioned} {Databases}},
	isbn = {978-3-540-28628-8},
	doi = {10.1007/978-3-540-28628-8_32},
	language = {en},
	booktitle = {Advances in {Cryptology} – {CRYPTO} 2004},
	publisher = {Springer},
	author = {Dwork, Cynthia and Nissim, Kobbi},
	editor = {Franklin, Matt},
	year = {2004},
	pages = {528--544},
}

@inproceedings{dinur_revealing_2003,
	address = {New York, NY, USA},
	series = {{PODS} '03},
	title = {Revealing information while preserving privacy},
	isbn = {978-1-58113-670-8},
	url = {https://doi.org/10.1145/773153.773173},
	doi = {10.1145/773153.773173},
	urldate = {2023-03-16},
	booktitle = {Proceedings of the twenty-second {ACM} {SIGMOD}-{SIGACT}-{SIGART} symposium on {Principles} of database systems},
	publisher = {Association for Computing Machinery},
	author = {Dinur, Irit and Nissim, Kobbi},
	month = jun,
	year = {2003},
	pages = {202--210},
}

@misc{mcmahan_communication-efficient_2023,
	title = {Communication-{Efficient} {Learning} of {Deep} {Networks} from {Decentralized} {Data}},
	url = {http://arxiv.org/abs/1602.05629},
	doi = {10.48550/arXiv.1602.05629},
	urldate = {2023-07-22},
	publisher = {arXiv},
	author = {McMahan, H. Brendan and Moore, Eider and Ramage, Daniel and Hampson, Seth and Arcas, Blaise Agüera y},
	month = jan,
	year = {2023},
	note = {arXiv:1602.05629 [cs]},
}

@misc{thapa_splitfed_2022,
	title = {{SplitFed}: {When} {Federated} {Learning} {Meets} {Split} {Learning}},
	shorttitle = {{SplitFed}},
	url = {http://arxiv.org/abs/2004.12088},
	urldate = {2023-01-20},
	publisher = {arXiv},
	author = {Thapa, Chandra and Chamikara, M. A. P. and Camtepe, Seyit and Sun, Lichao},
	month = feb,
	year = {2022},
	note = {arXiv:2004.12088 [cs]},
}

@article{zhang_survey_2021,
	title = {A survey on federated learning},
	volume = {216},
	issn = {0950-7051},
	url = {https://www.sciencedirect.com/science/article/pii/S0950705121000381},
	doi = {10.1016/j.knosys.2021.106775},
	language = {en},
	urldate = {2023-01-15},
	journal = {Knowledge-Based Systems},
	author = {Zhang, Chen and Xie, Yu and Bai, Hang and Yu, Bin and Li, Weihong and Gao, Yuan},
	month = mar,
	year = {2021},
	pages = {106775},
}

@misc{chen_targeted_2017,
	title = {Targeted {Backdoor} {Attacks} on {Deep} {Learning} {Systems} {Using} {Data} {Poisoning}},
	url = {http://arxiv.org/abs/1712.05526},
	doi = {10.48550/arXiv.1712.05526},
	urldate = {2022-12-13},
	publisher = {arXiv},
	author = {Chen, Xinyun and Liu, Chang and Li, Bo and Lu, Kimberly and Song, Dawn},
	month = dec,
	year = {2017},
	note = {arXiv:1712.05526 [cs]},
}

@misc{tolpegin_data_2020,
	title = {Data {Poisoning} {Attacks} {Against} {Federated} {Learning} {Systems}},
	url = {http://arxiv.org/abs/2007.08432},
	doi = {10.48550/arXiv.2007.08432},
	urldate = {2022-12-13},
	publisher = {arXiv},
	author = {Tolpegin, Vale and Truex, Stacey and Gursoy, Mehmet Emre and Liu, Ling},
	month = aug,
	year = {2020},
	note = {arXiv:2007.08432 [cs, stat]},
}

@misc{bagdasaryan_how_2019,
	title = {How {To} {Backdoor} {Federated} {Learning}},
	url = {http://arxiv.org/abs/1807.00459},
	doi = {10.48550/arXiv.1807.00459},
	urldate = {2022-12-13},
	publisher = {arXiv},
	author = {Bagdasaryan, Eugene and Veit, Andreas and Hua, Yiqing and Estrin, Deborah and Shmatikov, Vitaly},
	month = aug,
	year = {2019},
	note = {arXiv:1807.00459 [cs]},
}

@misc{sun_can_2019,
	title = {Can {You} {Really} {Backdoor} {Federated} {Learning}?},
	url = {http://arxiv.org/abs/1911.07963},
	doi = {10.48550/arXiv.1911.07963},
	urldate = {2022-12-09},
	publisher = {arXiv},
	author = {Sun, Ziteng and Kairouz, Peter and Suresh, Ananda Theertha and McMahan, H. Brendan},
	month = dec,
	year = {2019},
	note = {arXiv:1911.07963 [cs, stat]},
}

@misc{chen_federated_2022,
	title = {Federated {Learning} {Attacks} and {Defenses}: {A} {Survey}},
	shorttitle = {Federated {Learning} {Attacks} and {Defenses}},
	url = {http://arxiv.org/abs/2211.14952},
	doi = {10.48550/arXiv.2211.14952},
	urldate = {2022-12-08},
	publisher = {arXiv},
	author = {Chen, Yao and Gui, Yijie and Lin, Hong and Gan, Wensheng and Wu, Yongdong},
	month = nov,
	year = {2022},
	note = {arXiv:2211.14952 [cs]},
}

@article{liu_distributed_2022,
	title = {From distributed machine learning to federated learning: a survey},
	volume = {64},
	issn = {0219-3116},
	shorttitle = {From distributed machine learning to federated learning},
	url = {https://doi.org/10.1007/s10115-022-01664-x},
	doi = {10.1007/s10115-022-01664-x},
	language = {en},
	number = {4},
	urldate = {2022-12-08},
	journal = {Knowledge and Information Systems},
	author = {Liu, Ji and Huang, Jizhou and Zhou, Yang and Li, Xuhong and Ji, Shilei and Xiong, Haoyi and Dou, Dejing},
	month = apr,
	year = {2022},
	pages = {885--917},
}

\section{Appendix}
\subsection{Experiment results for other defense algorithms}

In this appendix we will present the relevant experiments for the rest of the defense algorithms, that did not produce the expected results when it comes to either accuracy in attacker detection, or to quantitative metrics for the performance of the model. Due to the poor performance of some of those algorithms, we will only present their results when trained on the MNIST dataset.

\subsubsection{Fixed Percentage Algorithm Results.}

We observe that he Source Class Recall metric performs very similarly to the honest model for even 40\% of the users being malicious. Those high numbers of the Source Class Recall are a clear indication that our defense mechanism performs as expected and eliminates malicious users from training. This was a given for small percentages of the users being malicious, but the fact that it performs the same way and produces similar results for high percentage of poisoning proves that when a fixed threshold is applied, the vast majority of malicious users are detected.  

\begin{figure}[h]
    \centering
    \includegraphics[width=.7\linewidth]{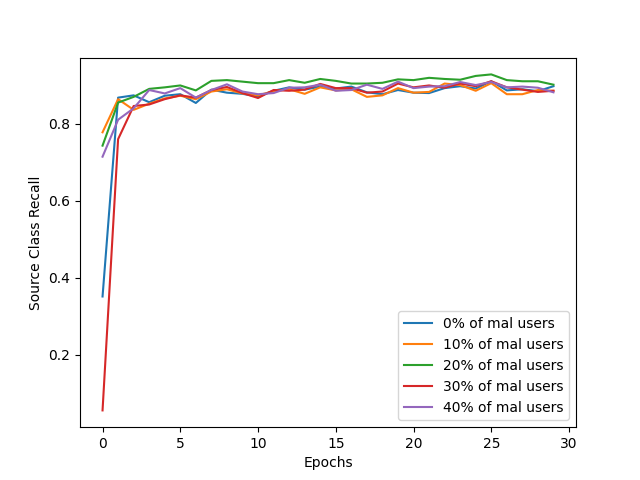}
    \caption{Source Class Recall for different percentages of the datasets being poisoned with the Fixed Percentage Algorithm present}
\end{figure}

\begin{figure}[h]
    \centering
    \includegraphics[width=.7\linewidth]{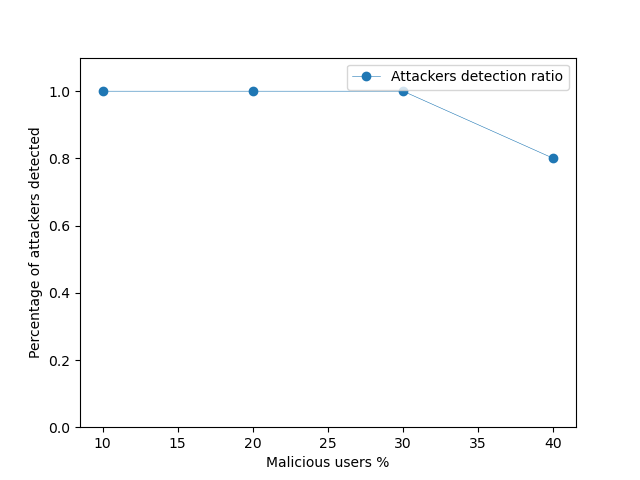}
    \caption{Attackers detection ratio by the Fixed Percentage Algorithm}
\end{figure}

Additionally, when it comes to attacker detection, we can observe \emph{Figure 12} that indeed the defense algorithm introduces a great increase in this metric in comparison with the original poisoning attack, for every one of our experiments with increasing malicious users’ percentages. 

To conclude, this method has some advantages but does not lack disadvantages. Without prior knowledge of the threshold of users that must be eliminated, that is almost impossible in a real-world scenario, the algorithm does not balance well in the scale of security and utility of the model. We saw that for low percentages of poisoning, when the threshold is high, it does not manage to produce good enough accuracy for the global FL model. Moreover it is safe to assume that for high percentages of malicious users, a low threshold would fail to detect all of them and would therefore not solve the problem that we are trying to defend against.

\subsubsection{Largest Difference Algorithm Results.}

In the relevant graphs, we can observe the failure of the algorithm with regards to the Source Class Recall metric, which reports low results for 40\% of the users being attackers. It indeed provides us with better curves for less attackers, something that is a good indicator for the performance of our algorithm for a reasonable number of attackers. 

\begin{figure}[t]
    \centering
    \includegraphics[width=.7\linewidth]{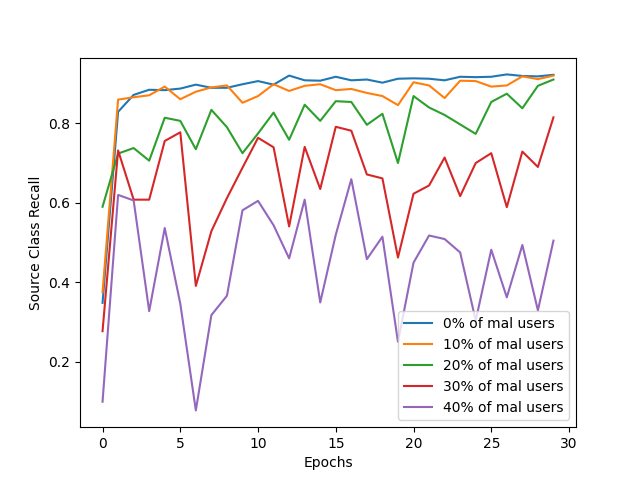}
    \caption{Source Class Recall for different percentages of the datasets being poisoned with the Largest Difference Algorithm present}
\end{figure}

\begin{figure}[t]
    \centering
    \includegraphics[width=.7\linewidth]{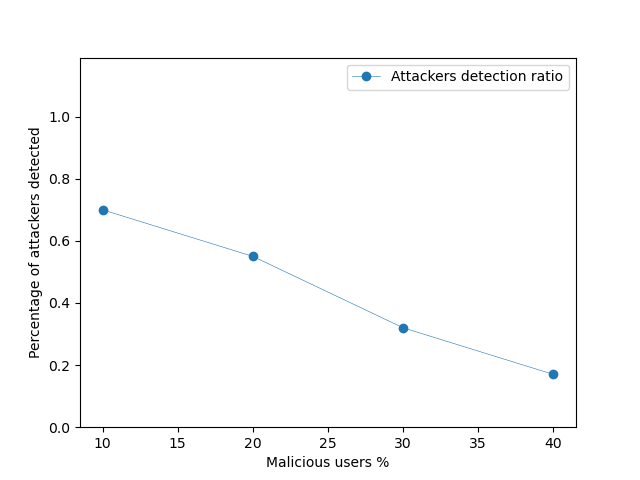}
    \caption{Attackers detection ratio by the Largest Difference Algorithm}
\end{figure}

The attackers are well enough spotted, as we can see in Figure 12, but decline as the percentage of them gets higher, something that checks out with the results that we presented earlier.

Thus, the Largest Difference algorithm seems to function as expected for low percentages of malicious clients, while failing to detect the majority of the users when this percentage rises to 40\%. When it comes to implementing in real-world application, the algorithm has the advantage of not requiring any prior knowledge, as the only arguments given are the losses and the total number of training clients, that is already known to the model. Moreover, in most cases it is extreme to assume that almost half of the clients selected each round for training will be malicious, hence this algorithm could prove useful for cases with less attackers, given its simplicity and its speed, as it as a linear complexity and does not introduce an extra computational overhead.

\subsubsection{Z-Score Algorithm Results.}

When taking a look at the Source Class Recall metric for this algorithm, we can confirm the good response to our training when the defense technique based on eliminating the users using this statistical measure, as the Recall after 30 rounds of training converges to high standards, above 0.8, even for 30\% of the users being malicious. For higher percentages, the results are not what we would like, however they were expected as we mentioned in the above theoretical foundations of the algorithm. 

\begin{figure}[t]
    \centering
    \includegraphics[width=.7\linewidth]{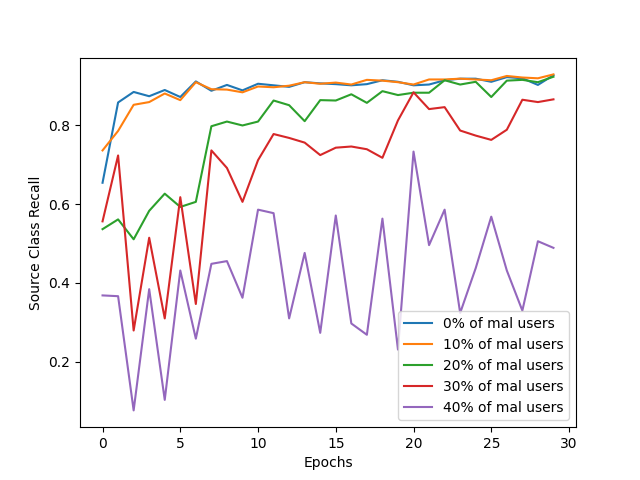}
    \caption{Source Class Recall for different percentages of the datasets being poisoned with the the Z-Score Algorithm present}
\end{figure}

\begin{figure}[t]
    \centering
    \includegraphics[width=.7\linewidth]{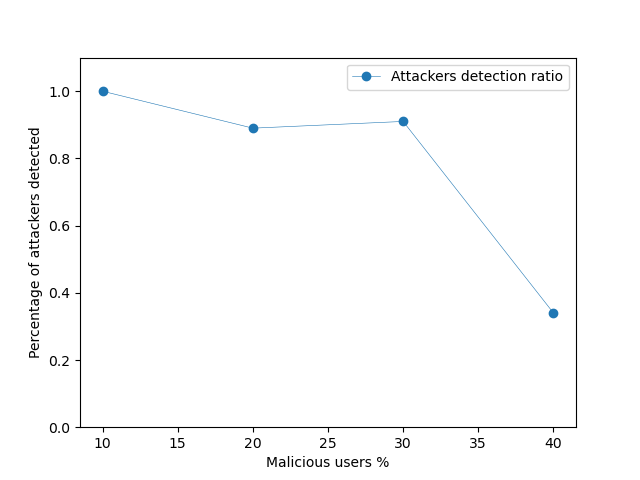}
    \caption{Attackers detection ratio by the Z-Score Algorithm}
\end{figure}

The above \emph{Figure 14} also confirms the correct classification of attackers, as we can see that their detection rate is high for the 3 first experiments, while it fails to keep up those numbers for higher percentages where it falls below 0.4.

To conclude, this method is deemed as extremely useful given the prior knowledge that there is a reasonable amount of poisoning in the model. In most real-world applications, it is safe to assume that 30\% or less of the users will be malicious, which makes this algorithm excellent to use, as it does not pose an extra computational overhead and can successfully classify attackers as such.

All the above algorithms presented, despite having their own scenarios where they perform well, they do not produce similarly good results for every case, something that is possible with the K-means clustering algorithm. Because of this, and due to the similar computational complexity and overhead, we conclude in promoting the K-means clustering algorithm for the task of classifying malicious users as such.


\end{document}